\documentclass[10pt,onecolumn,superscriptaddress,amsmath,amssymb,enumerate]{revtex4}
\usepackage{graphicx}
\usepackage{dcolumn}
\usepackage{bm}
\usepackage{color}
\usepackage{url}
\usepackage{tabularx}
\usepackage{array}
\usepackage{booktabs}
\usepackage{comment}
\usepackage{hyperref}
\usepackage{footnote}
\usepackage[normalem]{ulem}
\usepackage{braket}

\newcommand{\Ham}{\mathcal{H}}					
\newcommand{\beq}{\begin{equation}}				
\newcommand{\eeq}{\end{equation}}					

\begin{document}

\title{Interlayer exchange coupling -- a general scheme turning chiral magnets into magnetic multilayers carrying atomic-scale skyrmions}

\author{Ashis Kumar Nandy}\email{a.nandy@fz-juelich.de} 
\affiliation{Peter Gr\"unberg Institute and Institute for Advanced Simulation, Forschungszentrum J\"ulich and JARA, D-52425 J\"ulich, Germany}

\author{Nikolai S. Kiselev}
\affiliation{Peter Gr\"unberg Institute and Institute for Advanced Simulation, Forschungszentrum J\"ulich and JARA, D-52425 J\"ulich, Germany}

\author{Stefan Bl\"ugel} 
\affiliation{Peter Gr\"unberg Institute and Institute for Advanced Simulation, Forschungszentrum J\"ulich and JARA, D-52425 J\"ulich, Germany}

\begin{abstract}

We report on a general principle using the interlayer exchange coupling to extend the regime of chiral magnetic films in which stable or metastable magnetic skyrmions  can appear at zero magnetic field. We verify this concept on the basis of a first-principles model for a Mn monolayer on W(001) substrate, a prototype chiral magnet for which the atomic-scale magnetic texture is determined by the frustration of exchange interactions, impossible to unwind by laboratory magnetic fields. By means of \textit{ab initio} calculations for the Mn/W$_m$/Co$_n$/Pt/W(001) multilayer system we show that for certain thicknesses $m$ of the W spacer and $n$ of the Co reference layer, the effective field of the reference layer fully substitutes the required magnetic field for skyrmion formation.

\end{abstract}

\pacs{
      74.25.Ha   
      75.70.-i,  
      75.10.Hk,  
      31.15.A- 	 
      }

\maketitle

Chiral magnetic skyrmions are localized magnetic vortices with particle like properties.
They may occur as stable or metastable states in chiral magnets due to the competition between Heisenberg exchange and Dzyaloshinskii-Moriya interaction (DMI) \cite{Dzyaloshinckii, Moriya}, which is the essential ingredient for skyrmion stabilization.
Unique static and dynamic properties of magnetic skyrmions driven by their nontrivial topology make them attractive for the practical application in spintronic devices \cite{Fert_13,Kiselev_11} and interesting objects for fundamental research.

The DMI is the result of spin-orbit coupling (SOC) that occurs in magnetic systems with broken inversion symmetry.
The systems where chiral skyrmions have been observed so far can be divided into three main classes: (i) noncentrosymmetric bulk crystals, \textit{e.g.}\ MnSi \cite{Muehlbauer_09}, FeGe \cite{Wilhelm_11}, FeCoSi \cite{Munzer_10}, (ii) thin films of noncentrosymmetric crystals \textit{e.g.}\ MnSi \cite{Tonomura_12}, FeGe \cite{Yu_11}, FeCoSi \cite{Yu_10}, and (iii) ultrathin layers and multilayers with surface/interface induced DMI \cite{Heinze_11, Science_13, Bertrand_14}. 
The latter class of materials seems very promising, because they are most compatible to device technology and varying the thickness of the layers and the composition at the interface allow to tune the intrinsic parameters such as exchange, DMI, magnetic anisotropy  \textit{etc.}\ in a wide range \cite{Bertrand_14,Chen_13,Bertrand_arxiv}. 

Of particular interests are atomic-scale isolated skyrmions (iSk), whose stability is robust over a large range of magnetic fields and temperatures. 
This brings into play ultrathin layers of chiral magnets with non-micromagnetically describable magnetic behavior due to competing ferro- and antiferromagnetically coupled exchange interactions between different atomic sites that are finally the origin of stable achiral exchange spin-spirals (SS) of atomic length scale.
In this case, the role of the DMI is to select a particular chirality of spirals.
From micromagnetic theory \cite{Bogdanov_94&99, Melcher_14} it is known that for a given spin-stiffness, DMI and anisotropy, there is in principle always a range of applied magnetic fields that lead to the skyrmion phase.
We show that ultrathin films of chiral magnets with exchange driven SS posses the same properties as conventional chiral magnets, but the energy scale translates into gigantic magnetic fields that are not accessible in the laboratory, and thus those type of magnets had been excluded so far. 
A prototype system for this type of magnets is a monolayer (ML) of Mn  on a W(001) substrate with an experimentally confirmed~\cite{Ferriani_08} atomic-scale chiral exchange stabilized SS, but no skyrmions could be found.  

\begin{figure*}[!ht]
 \centering
   \includegraphics[width=17cm]{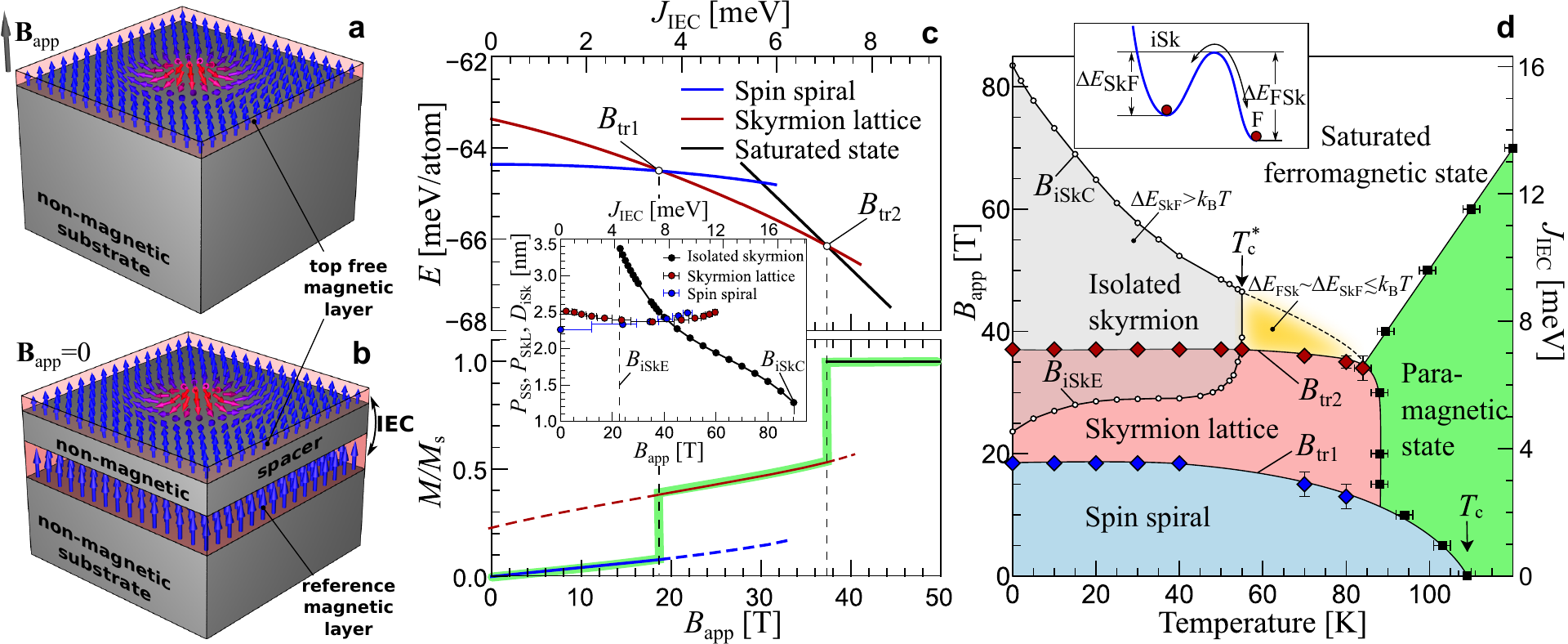}
\caption{(color online) Schematic representation of (a) the single magnetic layer system with skyrmions stabilized by a magnetic field, $B_\textrm{app}$, applied normally to the film and (b) an equivalent multilayer system with the $B_\textrm{app}$ replaced by the IEC between top free and underlying reference magnetic layers with a fixed out-of-plane magnetization.
(c) Energy density (top panel) and average out-of-plane magnetization (bottom panel) for spin-spiral (SS), hexagonal skyrmion lattice (SkL) and saturated ferromagnetic (FM) state as functions of $B_\textrm{app}$ (bottom axis) as well as the corresponding $J_\textrm{IEC}$ (top axis) calculated for Mn/W(001) at zero temperature. 
The solid green line corresponds to the equilibrium magnetization, dashed lines correspond to metastable states.
The inset shows the dependence of the equilibrium period length $P$ of SS, hexagonal SkL and the size of isolated skyrmion (iSk) as function of $B_\textrm{app}$ and $J_\textrm{IEC}$.
(d) Magnetic phase diagram for Mn/W(001). 
The SkL is energetically most favorable in the range between $B_\textrm{tr1}$ and $B_\textrm{tr2}$ (red area).
The range of existence for iSk (shaded gray area) is bounded by an elliptical instability field $B_\textrm{iSkE}$ and a collapse field $B_\textrm{iSkC}$ and intersects with the range of equilibrium SkL. 
The temperature stability of iSks is restricted by the critical temperature $T^{*}_\textrm{c}$, above which iSks appear in \textit{skyrmion soup} state (yellow area) where skyrmions with a short life time exhibit spontaneous annihilation and nucleation.
In the inset, $\Delta E_\textrm{SkF}$ and $\Delta E_\textrm{FSk}$ define the energy barriers for the transition from the iSk state to the FM state and the reverse one, respectively. 
For details see \cite{Supplementary_Materials}.}
\label{fig_MC}
\end{figure*}

In this Letter, we propose the interlayer exchange coupling (IEC) \cite{Grunberg_prl,Parkin_prl,Bruno_IEC1,Bruno_IEC2} as an elegant approach to stabilize skyrmions without external field and thus widen the class of chiral magnets hosting skyrmions. 
We justify our concept  by  calculations within a multiscale model based on \textit{ab initio} calculations and atomistic simulations in the frame of a classical spin model.
The model put forward applies to ultrathin transition-metal films deposited on a nonmagnetic heavy metal substrate where the DMI is induced by the interface due to the strong SOC \cite{Bode_Nat}. 
In such a system, the skyrmions may appear under a magnetic field applied perpendicular to the layer, Fig.~\ref{fig_MC}a. 
Alternatively, we propose that the skyrmion phase can be stabilized by designing multilayers composed of two magnetic layers (top free and bottom reference layers) separated by a nonmagnetic spacer layer, where the applied magnetic field is substituted by the IEC, which acts as an effective magnetic field, see Fig.~\ref{fig_MC}b. 
In this approach, the system should satisfy the following conditions: (i) the reference layer should be a hard ferromagnet with large exchange stiffness and strong out-of-plane anisotropy, (ii) the effective field induced by IEC should be in the range of the magnetic field required for skyrmions stabilization and (iii) the internal properties, mainly the exchange coupling and the strength of DMI of the free layer remain nearly the same as in the parent system.

To describe such a multilayer system we use the following model Hamiltonian, comprising the contribution from Heisenberg exchange, DMI, out-of-plane anisotropy, Zeeman and IEC between free and reference layer interactions: 
\begin{align}
\Ham  =&   -\! \sum_{i<j}\! J_{ij} (\hat{m}_i \!\cdot\! \hat{m}_j) 
                            -\!\! \sum_{\left\langle i<j \right\rangle}\!\! \vec{D}_{ij}\! \cdot\! [\hat{m}_i\! \times\! \hat{m}_j]  -\! \sum_{i}\! K (m^z_{i})^2 \nonumber \\
                          & -\! \sum_{i}\!\mu \vec{B}_\textrm{app} \!\cdot\! \hat{m}_i -\!\sum_{\left\langle i<j \right\rangle}\!{J_\textrm{IEC}(\hat{m}_{i}\cdot \hat{m}_{\textrm{R}j})}, 
\label{eq_modelHamiltonian}
\end{align}
where $\left\langle i<j \right\rangle$ denotes summation over all the nearest-neighbor pairs.
$\hat{m}$ and $\hat{m}_\textrm{R}$ are unit vectors of magnetic moment in the free and reference layers, respectively and $\mu$ is the absolute value of the magnetic moment of a free layer atom.
%
%
For a fixed magnetization in the hard reference layer, $\vec{m}_{\textrm{R}i} = \vec{m}_\textrm{R}$ for all sites $i$, the IEC term ${\cal H}_\textrm{IEC}$ can be rewritten as ${\cal H}_\text{IEC} = -\sum_{i}{\mu \vec{B}_\textrm{eff}\cdot \hat{m}_{i}}$ with
\begin{equation}
\vec{B}_\textrm{eff} = J_\textrm{IEC}\hat{m}_\textrm{R}/\mu.
\label{eq_iec2}
\end{equation}  
Thereby, in case of out-of-plane anisotropy, $\vec{B}_\textrm{eff}$ always points either parallel or antiparallel to the multilayer depending on the sign of $J_\textrm{IEC}$.
The $J_\textrm{IEC}$ is determined form first-principles as half of the total-energy difference between the parallel and antiparallel orientation of the magnetic moments of the free and reference layers.
 
We apply this model to the prototype system Mn/W(001) with parameters including Heisenberg exchange up to 7\textsuperscript{th} neighbor shells, nearest-neighbor DMI and uniaxial anisotropy, all obtained from first-principles \cite{Supplementary_Materials}. 
We estimated the contribution of the dipole-dipole interactions and found its contribution negligibly small with respect to other energy terms \cite{Supplementary_Materials}.
We confirm an atomic-scale chiral SS ground state in excellent agreement with experiment \cite{Ferriani_08}. 

We have minimized the Hamiltonian (\ref{eq_modelHamiltonian}) considering two equivalent limiting cases, $B_\textrm{app}=0$ and $J_\textrm{IEC}=0$. 
We have identified the critical fields $B_\textrm{tr1}$ and $B_\textrm{tr2}$ and the corresponding IEC for the transition between the SS, hexagonal skyrmion lattice (SkL) and saturated FM states, see Fig.~\ref{fig_MC}c. 
To emphasize the equivalence of the two limiting cases, both values of the $B_\textrm{app}$ and the $J_\textrm{IEC}$ are given in the bottom and top axes in Fig.~\ref{fig_MC}c, respectively.
The lower panel shows the equilibrium magnetization accompanied with the jumps at corresponding phase transitions.
The inset shows the equilibrium period lengths of the SS ($P_\textrm{SS}$) and the hexagonal SkL ($P_\textrm{SkL}$) and the diameter ($D_\textrm{iSk}$) of an iSk.
Obviously, we expect atomic-scale skyrmions with a diameter of 2-3 nm.

Fig.~\ref{fig_MC}c illustrates the significant difference between SkL and iSk. 
Contrary to the SkL which may appear as a metastable state at zero magnetic field, the iSk at low field exhibits an elliptical instability at $B_\textrm{iSkE}$ and does not exist for fields below $B_\textrm{iSkE}$. 
Moreover, the iSk shows remarkable size variations in strong contrast to the small changes in the period length of the SkL. 
Note, in the case of very strong uniaxial anisotropy metastable iSk can be stabilized at zero applied field \cite{Bogdanov_94&99}.
However, the size of such skyrmions and required values of anisotropy became unrealistic, for detail see \cite{Supplementary_Materials}.

To illustrate the temperature dependence of critical fields and the corresponding IEC, using Monte-Carlo simulations (MCS) we have calculated the magnetic phase diagram presented in Fig.~\ref{fig_MC}d and aslo in \cite{Supplementary_Materials}. 
%
This phase diagram describes the general behavior of two-dimensional chiral magnets and is in a good qualitative agreement with experimental observations \cite{Yu_10,Yu_11} as well as with recently reported results of Monte Carlo simulations for Pd/Fe/Ir \cite{Rozsa_PRB}.
Both the phase transition lines between SS and SkL as well as between SkL and FM states exhibit only a weak temperature dependence and therefore, the applied magnetic field or the IEC required to stabilize the SkL remains the same even at high temperatures. 
On the other hand, the range of existence for iSk strongly depends on the temperature.
%
We have found a critical temperature $T^{*}_\textrm{c}$ above which the average energy of the thermal fluctuations becomes higher than the energy barrier which protects the iSk from the collapse.
In this particular case of Mn/W(001), $T^{*}_\textrm{c}$ is found to be approximately half the ordering temperature $T_\textrm{c}\approx110$~K, but in general, it is a function of material parameters and may vary for different systems.
In a certain region above $T^{*}_\textrm{c}$, marked as yellow area in the phase diagram, spontaneous annihilation and nucleation of iSk takes place. 
We refer to this state as the boiling \textit{skyrmion soup}, because skyrmions appear and disappear as bubbles on the surface of the boiling water.
The physical reason for the appearance of the \textit{skyrmion soup} is that the energy barriers for the transition between iSk and FM state, $\Delta E_\textrm{SkF}$, as well as the reverse one, $\Delta E_\textrm{FSk}$, become comparable to the energy of thermal fluctuations $k_\textrm{B}T$, see inset in phase diagram.
The \textit{skyrmion soup} state also can be interpreted as the special B-T range where the saturated ferromagnetic ground state becomes unstable with respect to spontaneous nucleation and following annihilation of the skyrmions.
Experimental measurements for the skyrmion life-time in this state provide direct access to the estimation of the energy barriers controlling skyrmion nucleation and annihilation processes.
As far as the stability range of iSk is temperature dependent, it is important to identify the optimal magnetic field for which the iSk remains stable within the whole range of temperature between 0~K and $T^{*}_\textrm{c}$. 

According to the phase diagram, both phases iSk and SkL are stable over a large range of magnetic fields and temperature. The optimal field to stabilize iSk has to be fixed slightly above $B_\textrm{tr2}$. For the SkL, this field has to be between $B_\textrm{tr1}$ and $B_\textrm{tr2}$, both can be experimentally identified from the jumps on the magnetization curve, Fig.~\ref{fig_MC}c.
For the prototype system Mn/W(001) discussed in this letter, we indeed find magnetic fields of gigantic values, $B_\textrm{tr1}\approx18$~T and $B_\textrm{tr2}\approx37$~T.

\begin{figure}[!t]
 \centering
 \includegraphics[width=10cm]{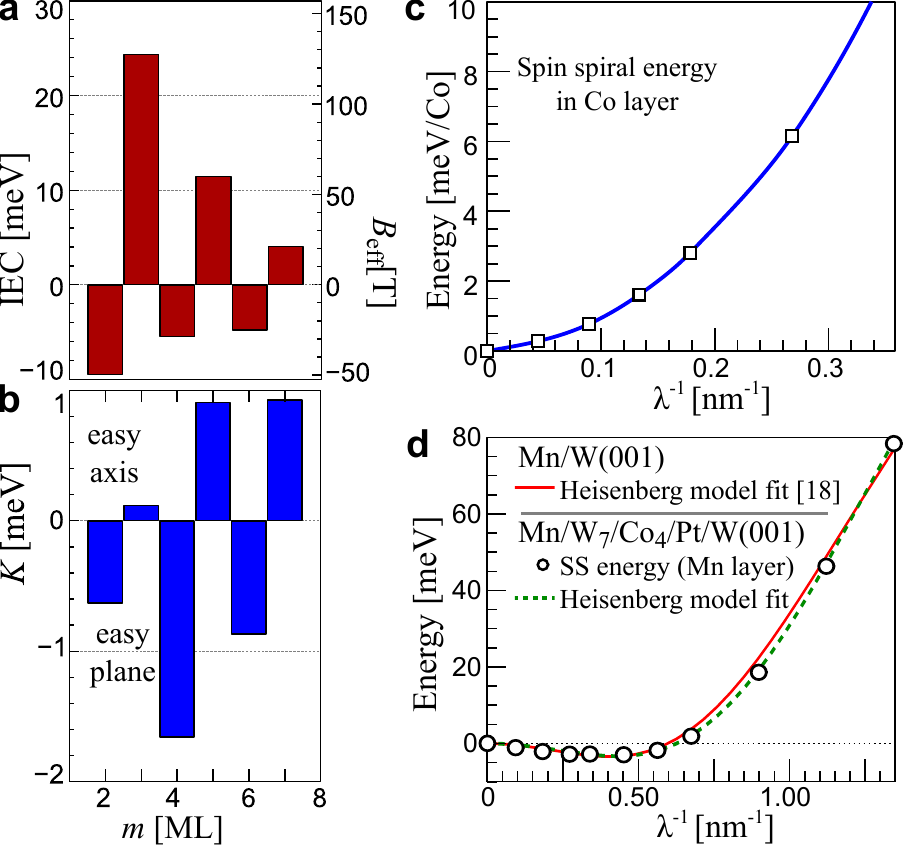}
 \caption{(color online) (a) Strength of the IEC between Mn and Co and (b) MCA of Co monolayer with respect to the thickness \textit{m} of the W spacer, corresponding to the calculation for Mn/W$_m$/Co$_n$/W(001) with $n=1$.
For Mn/W$_7$/Co$_4$/Pt/W(001), (c) and (d) represent the energy of a flat homogeneous SS for the reference Co (squares) and the free Mn (circles) layers, respectively, calculated without SOC with respect to inverse SS period length, $\lambda^{-1}$, along the $\Braket{110}$ direction.
The dashed (green) line is the fit to the Heisenberg model to calculate the exchange parameters.
Solid (red) line is a fit to the Heisenberg model for the pristine Mn/W(001) system, used for the calculation presented in Fig.~\ref{fig_MC}.}
 \label{fig_ab-initio}
\end{figure}

In order to realize the appropriate field $B_\textrm{eff}$ in terms of the IEC field (Eq.~\ref{eq_iec2}) exerted on the Mn free layer of the Mn/W(001) system, we deviced a realistic multilayer system, Mn/W$_m$/Co$_n$/W(001) and estimated the number of atomic-layers $m$ and $n$  by  \textit{ab initio} calculations to design a reference layer that fulfills conditions (i)$-$(iii) mentioned above.
For the case of $n$=1, the results of the IEC between Mn and Co and the magnetic anisotropy of Co as function of the  number of W spacer layers $m$  are presented in Fig.~\ref{fig_ab-initio}a and b, respectively.
The IEC in Fig.~\ref{fig_ab-initio}a exhibits an oscillatory behavior and for $m\leq7$ the corresponding effective field varies in the range of few tens of Tesla.
For the spacer thickness $m=5$, 6, and 7, the effective magnetic fields are about 60, 25, and 21~T, respectively, which are inside the range of magnetic fields required for skyrmion stabilization, see Fig.~\ref{fig_MC}c and d. 
The thickness of the nonmagnetic spacer affects also the magnetocrystalline anisotropy (MCA) energy, Fig.~\ref{fig_ab-initio}b. 
The magnetic moments of about $\mu_\textrm{Mn}=3.1~\mu_\textrm{B}$ and $\mu_\textrm{Co}=1~\mu_\textrm{B}$ on Mn and Co, respectively, depend only weekly on $m$ (not shown).

\begin{table}[!b]
\begin{center}
\caption{Results of \textit{ab initio} calculations for multilayers of different geometry, Mn/W$_7$/Co$_n$/Pt/W(001).
Magnetic moment of Co, $\mu_\textrm{Co}$ as well as energy of FM stability, $\Delta\epsilon$, and MCA, $\tilde{K}_\textrm{Co}$, of the reference layer calculated per magnetic Co atom.
In particular, for the last example system, the Co magnetic moment at the W/Co interface is very small compared to other Co atoms and we consider it as a nonmagnetic atom in Eq.~\ref{eq_modelHamiltonian}. 
$\tilde{K}_\textrm{Co}>0$ refers to out-of-plane easy axis.}
\label{tab_robust-FM}
\begin{tabular}{lccc}\hline\hline
Multilayer & Magnetic moment & $\Delta\epsilon$ & $\tilde{K}_\textrm{Co}$   \\ 
system & of Co ($\mu_\textrm{B}$) & (meV) & (meV) \\
\hline
Mn/W$_7$\!/Co$_1$\!/W(001) & 1.04 & \phantom{35}1 & \phantom{$-$}0.93 \\
Mn/W$_7$\!/Co$_3$\!/W(001) & 0.52,\! 1.29,\! 0.29 & \phantom{3}15 & $-$0.05 \\
Mn/W$_7$\!/Co$_3$\!/Pt/W(001) & 0.44,\! 1.56,\! 1.68 & \phantom{1}68 & $-$0.41 \\
Mn/W$_7$\!/Co$_4$\!/Pt/W(001) & 0.15,\! 1.57,\! 1.72,\! 1.62 & 135 & \phantom{$-$}1.12 \\
\hline\hline
\end{tabular}
\label{tab_robust-FM}
\end{center}
\end{table}

Another important parameter which needs to be examined is the exchange stiffness of the Co layer which has to be large enough to realize a hard FM reference layer.
In Table~\ref{tab_robust-FM}, we present the \textit{ab inito} results for multilayer systems of different Co thicknesses and interfaces for a given W spacer layer. 
The energy difference $\Delta\epsilon>0$ represents the stability energy of the FM state of Co calculated with respect to the antiferromagnetic $c(2\times2)$ state, which is known to be the ground state for the Co ML on W(001) system \cite{Ferriani-prb}.
The higher value of $\Delta\epsilon$ attributes to the higher exchange stiffness and $\Delta\epsilon$ increases with the number of Co layers. 
We have found that the Co/W interface significantly reduces the magnetic moment of the Co atom at the interface and inhibites the hardness of the thin reference layer.

To achieve a high exchange stiffness we modify the Co/W interface by introducing an additional Pt ML.
The Co/Pt/W interface shows a strong influence on $\Delta\epsilon$ and $\mu_\textrm{Co}$, both are increased significantly compared to the pure Co/W interface (compare rows two and three in Table~\ref{tab_robust-FM}). 
The thickness and the interface composition also affect the MCA of the Co layer (see the last column in Table~\ref{tab_robust-FM}). 
For the case of 4 Co MLs, both the strong out-of-plane $\tilde{K}_\textrm{Co}$ and the large positive $\Delta\epsilon$ are achieved.

Focusing on the Co$_4$/Pt/W(001) system, to characterize the Mn/W$_m$/Co$_4$/Pt/W(001) system, for $m=7$, we have calculated from first-principles the energy dispersion for homogeneous flat SSs in the reference and free layers, Fig.~\ref{fig_ab-initio}c-d.
Such a SS is characterized by a wave vector \textbf{q} or period length $\lambda=2\pi/|\textbf{q}|$.
Following the approach proposed in Ref.~\onlinecite{Bernd_14}, from fits to the energy dispersion we extracted the parameters $\tilde{A}_\textrm{Co}=A/V_\mathrm{Co}$, $\tilde{D}_\textrm{Co}=D/V_\mathrm{Co}$ and $\tilde{K}_\textrm{Co}=K/V_\mathrm{Co}$ where $A$, $D$ and $K$ are micromagnetic constants of exchange stiffness, DMI and uniaxial anisotropy, respectively, and $V_\mathrm{Co}$ is the average volume per single Co atom in the unit cell.
The interplay between these three quantities determines the magnetic ground state of a system.
The criterion for the stability of homogeneous FM state is $\kappa=\pi D/(4\sqrt{AK})<1$ \cite{helix}. 

An exchange stiffness constant of $\tilde{A}_\textrm{Co}=90$~meV$\cdot$nm$^2$ and an average DMI constant of $\tilde{D}_\textrm{Co}$=$-$1.62~meV$\cdot$nm are calculated from the energy dispersion without and with SOC, respectively (for details, see \cite{Supplementary_Materials}).
Note, both the interfaces W/Co ($-$1.9~meV$\cdot$nm) and Co/Pt ($-$2.7~meV$\cdot$nm) contribute to the average $\tilde{D}_\textrm{Co}$.
We estimate the average out-of-plane anisotropy of the Co layer to be $\tilde{K}_\textrm{Co}=1.12$~meV. 

Taking into account the values of $\tilde{A}_\textrm{Co}$, $\tilde{D}_\textrm{Co}$ and $\tilde{K}_\textrm{Co}$ and their ratio to corresponding micromagnetic constants, we have estimated $\kappa=0.13$ which shows that the assumption of a hard FM reference layer with out-of-plane anisotropy as required in our model is fully satisfied for the prototype system of Mn/W$_7$/Co$_4$/Pt/W(001).

Finally, we examine the influence of the underlying Co reference layer on the coupling parameters of the Mn free layer.
From the \textit{ab initio} calculations for Mn/W$_7$/Co$_4$/Pt/W(001) we have estimated the coupling constants for the Heisenberg exchange as the dominant energy term.
In Fig.~\ref{fig_ab-initio}d, the energy dispersion of the flat SS in the Mn layer is shown. 
One can see an equivalent behavior to the pristine Mn/W(001) system.
A fit to the Heisenberg model reveals that the exchange interactions $J_{ij}$ remain almost unchanged, see \cite{Supplementary_Materials}.
The DMI, which is the sum of the layer resolved contributions of the first few layers~\cite{Bernd_14} remains unchanged for a  nonmagnetic spacer thick enough, $m\ge5$.
For Mn/W$_7$/Co$_4$/Pt/W(001), we have found an about 56\% higher out-of-plane MCA, $\tilde{K}_\textrm{Mn}=5.6$~meV of the Mn layer as compared to the pristine Mn/W(001), $3.6$~meV.  

Summarizing all results for the Mn/W$_m$/Co$_n$/Pt/ W(001) multilayer system, we conclude that the multilayers engineered with $m=5$ and 6 W spacer layers and with a reference layer of $n=4$ Co layers are the best candidates for stabilizing iSk and SkL, respectively, without applied magnetic field.
In the systems with $m=5$, 6, and 7 layers of W, the IEC exerts huge effective out-of-plane magnetic fields of about 42, 22, and 15~T onto Mn with a strong out-of-plane anisotropy $\tilde{K}_\textrm{Co}$ of about 0.95, 0.83, and 1.12~meV/Co atom, respectively. 
The presence of the reference layer modifies slightly the MCA energies of Mn, $\tilde{K}_\textrm{Mn}$, to about 4.4, 4.5, and 5.6~meV, respectively, but this influences a little on the skyrmion size and skyrmion formation.

In conclusion, we extended the micromagnetic concept of stabilizing skyrmions by applied magnetic fields to skyrmions stabilized by interlayer exchange coupling. 
This enables the skyrmion formation in chiral magnets with competing exchange interactions that lead to more complicated magnetic ground states such as exchange spin-spirals that result in atomic-scale skyrmions, which may be robust over wide temperature and magnetic field ranges.  
Replacing the prototype system Mn/W(001) by the multilayer system Mn/W$_m$/Co$_n$/Pt/W(001), we have shown that by varying the geometrical parameters such as thickness and fixed-layer compositions, one can achieve a stabilization of small-scale ($D_\textrm{iSk}\approx2$~nm) magnetic skyrmions even when the required applied magnetic field would have been gigantic.    
Our approach is rather general and can be used for any two-dimensional chiral magnet with surface/interface induced DMI and thus provides a perspective direction to extend the number of possible systems where magnetic skyrmions can be observed also at elevated temperatures.

\begin{acknowledgments}
The authors acknowledge assistance by David S. G. Bauer and P. Ferriani in evaluating the parameters entering the model Hamiltonian. 
The authors thank B. Dup\'e, B. Zimmermann, S. Heinze, A. Fert and V. Cros for fruitful discussions and acknowledge M. Hoffmann for critical reading of the manuscript.
The authors acknowledge financial support from the European Union (FET-Open MAGicSky n\textsuperscript{o}665095). Computations were performed under the auspices the J\"ulich Supercomputing Centre.
\end{acknowledgments}

\newpage
\part*{\Large \centering Supplementary Materials}







\maketitle

\renewcommand{\thesection}{\text{S}\arabic{section}}
\numberwithin{equation}{section}
\numberwithin{figure}{section}
\numberwithin{table}{section}

\setcounter{equation}{0}
\setcounter{figure}{0}

\section{Details of the \textit{ab initio} calculations}

All the first-principles results presented in the main text are performed using the FLEUR \cite{fleur} implementation of the all-electron full-potential linearized augmented plane-wave method \cite{FLAPW1} in the film geometry \cite{FLAPW2}.
Our asymmetric multilayer, Mn/W$_m$/Co$_n$/W(001), consists of an atomic-layer of Mn, different thicknesses of W spacer layers, a Co reference layer of different thicknesses and eight layers of W(001) as the substrate with the experimental lattice constant 3.165 \AA~of bulk W.
We checked the number of W layers used as a substrate and do not find any quantitative changes for higher number of W layers.
Most of the calculations are performed using a $p(1\times1)$ surface unit cell, which comprises only one atom per layer.
The inter-layer distances in each geometry is relaxed within the generalized-gradient approximation (GGA) using Perdew, Burke, and Ernzerhof \cite{pbe} functional.
The atomic structures are considered relaxed when all the forces on the atoms comprising the Mn/W interface, spacer, reference layer, W/Co, Co/W, Co/Pt, Pt/W interfaces as well as the first three sub-layers of the substrate close to the reference layer are smaller than 1mRyd/a.u.
Other quantities such as magnetocrystalline anisotropy (MCA), ferromagnetic stability of the reference Co layer $\Delta\epsilon$, spin-spiral (SS) energy have been calculated using the LDA functional \cite{lda}.
The convergence with the cutoff energy for plane wave expansion of wave-functions and $k$-points used for two-dimensional Brillouin-zone (BZ) integration has been checked carefully.
For collinear (spin-spiral) calculations we have used a wave vector $k_\textrm{max}$ = 3.8 a.u.$^{-1}$ (4.0 a.u.$^{-1}$) to expand the LAPW basis functions.
In most collinear calculations with the $p(1\times1)$ unit cell, the charge densities are converged using $\bf k_\|$ = 256 in the full Brillouin zone.
The antiferromagnetic state is constructed using the $c(2\times2)$ surface supercell of each film geometry with and without the top Mn layer. $\Delta\epsilon$ is estimated per magnetic Co atom.
For the MCA calculations, the spin-orbit interactions have been considered using the force theorem where a dense $\bf k_\|$-point mesh (2500) is used in the full BZ.
In order to calculate the Dzyaloshinskii-Moriya interaction (DMI), we have calculated the energy shift of each SS state due to the spin-orbit coupling (SOC) within first-order perturbation theory.
The calculations of flat SS corresponding to Co and Mn layer in Mn/W$_7$/Co$_4$/W(001) multilayer are carried out using 1600 and 2500 $\bf k_\|$-points, respectively, in the full BZ.
Note, individual SS calculations of top Mn and reference Co layers are carried out such that the magnetic moments of the layer for which the SS calculation is performed are confined in the plane while rest of the magnetic moments are considered perpendicular to the plane. 
Therefore, Co and Mn layers are magnetically decoupled during individual energy dispersion calculations.

\section{Ground state of Mn/W(001) as well as the Dzyaloshinskii-Moriya interaction and exchnage stiffness for the Co layer and Heisenberg exchange parameters for the Mn layer in Mn/W$_7$/Co$_4$/W(001)}

\begin{figure}[!ht]
\includegraphics[width=12.5cm]{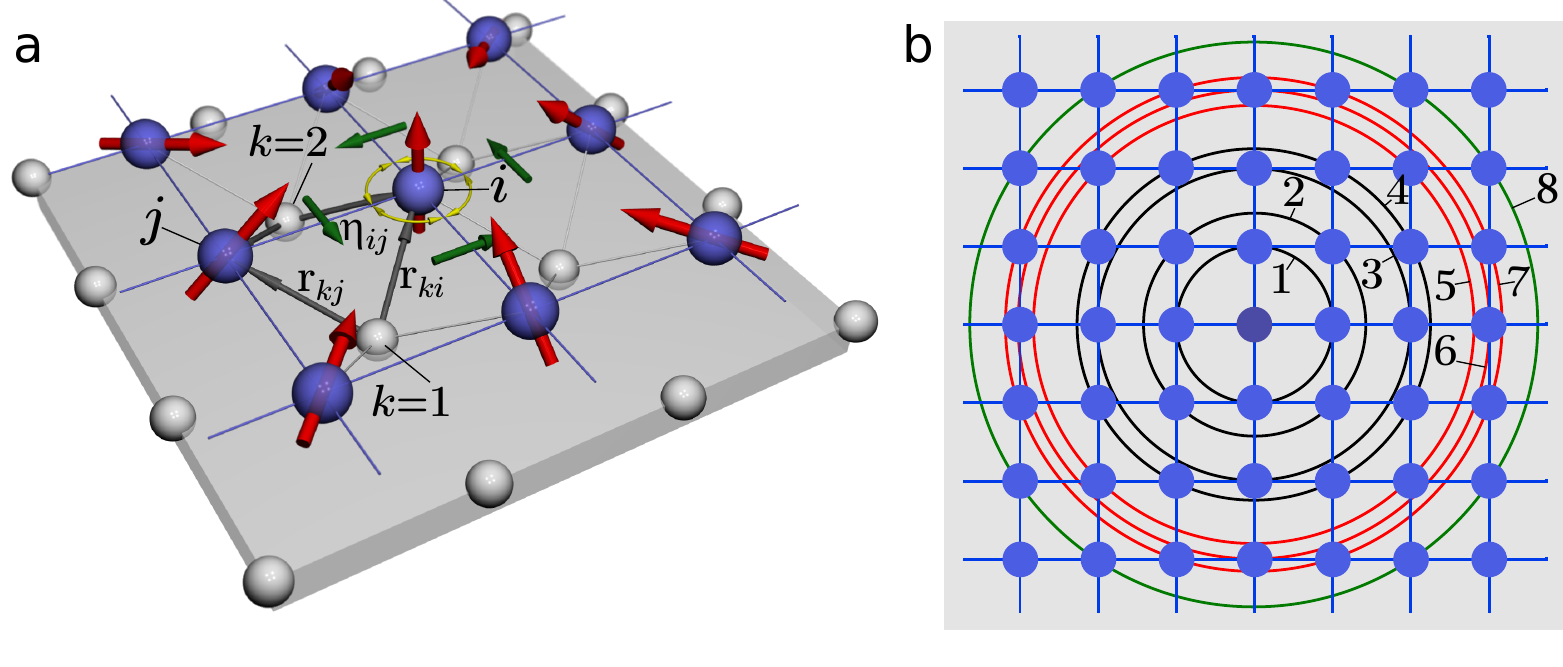}\\
 \caption{ Figure illustrates a fourfold-symmetric square lattice with the surface atoms Mn on top (blue spheres) and an adjacent layer of heavy atoms W (gray spheres). 
The vectors of magnetic moments of Mn atoms are marked as red arrows. The nearest-neighbor DM unit vectors between the central and the nearest-neighbor Mn atoms are shown as green arrows. 
The resultant unit vector $\boldsymbol\eta$ is inplane.}
\label{fig_DMI}
\end{figure}

\begin{figure}[!t]
\includegraphics[width=12.5cm]{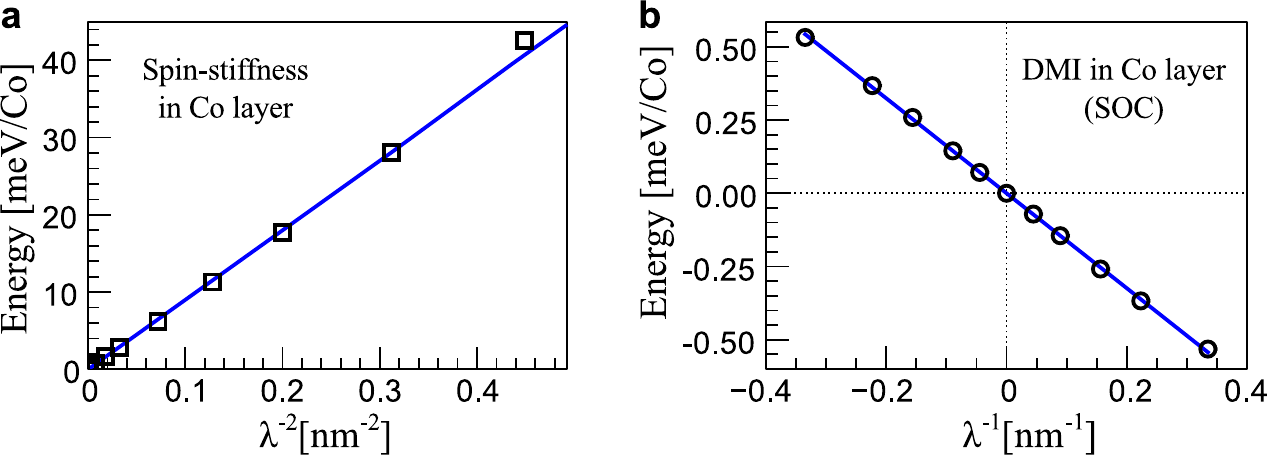}\\
 \caption{ Corresponding to the multilayer, Mn/W$_7$/Co$_4$/Pt/W(001), (a) the energy of a flat homogeneous SS (squares) for the reference Co layer calculated without SOC is shown with respect to $\lambda^{-2}$ and (b) the shift in energy of the SS (circles) due to the SOC within first order perturbation theory is shown with respect to $\lambda^{-1}$.
$\lambda$ is the period length of the SS.}
\label{fig_ab-initio_supple}
\end{figure}

In order to confirm the validity of our approach first we compare the theoretical and experimentally observed ground state of the prototype system Mn/W(001).
We adapted the set of $J_{ij}$'s given in Ref.~\onlinecite{Ferriani_08}. 
In comparison to Ref.~\onlinecite{Ferriani_08}, we improved the fitted values of the exchange parameters $J_{ij}$'s by noticing that in Ref.~\onlinecite{Ferriani_08} all $J$s up to the 5\textsuperscript{th} shell of neighboring atoms were fitted, but according to Fig.~\ref{fig_DMI}b, the distances of the atoms in the shells 5-7 are very similar.
We suppose a restriction of the fit up to the 5\textsuperscript{th} shell is inconsistent and we improve the fit by including also the atoms of shells 6 and 7. 
The final set of $J_{ij}$ used in the Monte Carlo simulations is provided in the first row of Table~\ref{tab_exchange}. 
We use only a nearest-neighbor Dzyaloshinskii-Moriya interaction (DMI) with the absolute value $D = -9.2$ meV per pair of Mn atoms. 
This is consistent with value of $+4.6$ meV per Mn atom of Ref.~\onlinecite{Ferriani_08}, due to a slightly different definition of the DMI.
Normalized direction of Dzyaloshinskii-Moriya vector $\boldsymbol\eta_{ij} = \frac{\textbf{D}_{ij}}{|\textbf{D}_{ij}|}$ is defined according to the Fert-Levy approach \cite{Fert&Levy} as
\begin{equation}
\boldsymbol\eta_{ij} = \frac{\sum_{k}{\mathbf{r}_{ki} \times \mathbf{r}_{kj}}}{|\sum_{k}{\mathbf{r}_{ki} \times \mathbf{r}_{kj}}|},
\label{eq_DM}
\end{equation}
where $\mathbf{r}_{ki}$ is the vector connecting positions of the nonmagnetic W atom $k$ and the magnetic Mn atom $i$. 
Note, the summation in equation~(\ref{eq_DM}) runs over the neighboring nonmagnetic atoms, which are mutual neighbors for both magnetic atoms $i$ and $j$. 
In this system, the DMI unit vector, $\boldsymbol\eta_{ij}$, defined by \ref{eq_DM} is found to be inplane between nearest-neighbor Mn atoms at the surface, see Fig.~\ref{fig_DMI}a, as can be implied by Moriya’s [2] (in the main text) third symmetry rule and consistent with the results of Udvardi \textit{et al.} \cite{Udvardi}.
For the magnetic anisotropy constant, $K = 3.6$~meV has been used. 

With this set of parameters we reproduced the ground state of the Mn/W(001) system with a high accuracy. 
The SS with period $P_\textrm{SS} \backsimeq$ 2.25 nm propagating along $\braket{110}$ directions has been found to be the ground state which is in a good agreement with experiment. 
In particular, spin-polarized scanning tunneling microscopy experiments on Mn/W(001) found the same $\braket{110}$ propagation direction of the SS and the period was estimated as $\sim$ 2.2 nm \cite{Ferriani_08}.

A linear fit (solid line) to the dispersion energies in Fig.~\ref{fig_ab-initio_supple}a results in an exchange stiffness constant of $\tilde{A}_\textrm{Co}=90$~meV$\cdot$nm$^2$. 
On the other hand, in Fig.~\ref{fig_ab-initio_supple}b, an average DMI constant of $\tilde{D}_\textrm{Co}$=$-$1.62~meV$\cdot$nm is calculated by a linear fitting (solid line) to the energy shifts due to the SOC.

In Table~\ref{tab_exchange}, the Heisenberg exchange interactions extracted for Mn spiral in the equivalent multilayer system Mn/W$_7$/Co$_4$/Pt/W(001) do not differ significantly to the corresponding exchange interactions used in the Monte Carlo (MC) simulations for the pristine Mn/W(001) system. 
Moreover, these parameters with the extracted magnetocrystalline anisotropy of Mn ($\approx$~5.6 meV) results in the same period length of SS as the ground state for the equivalent system.

\begin{table*}[!b]
\begin{center}
\caption{\label{tab_exchange} Exchange interactions extracted from Heisenberg model for Mn SS in Mn/W$_7$/Co$_4$/Pt/W(001) multilayer is compared with pure Mn/W(001) system used in our MC simulations.}
\begin{tabular}{lccccccc}\hline\hline
\multicolumn{1}{l} {Equivalent} & \multicolumn{7}{c} {Heisenberg exchange coupling in meV} \\ \cline{2-8}
systems & $J_1$ & $J_2$ & $J_3$ & $J_4$ & $J_5$ & $J_6$ & $J_7$  \\
\hline
Mn/W(001) & 19.7 & $-$2.5 & $-$1.5 & $-$0.5 & $-$0.56 & $-$0.56 & $-$0.28 \\
Mn/W$_7$\!/Co$_4$\!/Pt/W(001) & 20.25 & $-$2.9 & $-$1.81 & $-$0.45 & $-$0.15 & $-$1.2 & $-$0.03 \\
\hline\hline
\end{tabular}
\end{center}
\end{table*}

\section{The role of dipole-dipole interaction}
In this section we present a simple approach which allows to estimate the contribution of the dipole-dipole interaction (DDI). 
Here we made the estimation of the DDI contribution for two-dimensional (2D) systems, but in general this approach can be easily generalized also for three-dimensional case.

The general form for the energy of dipole-dipole interaction for atomistic models is 

\begin{equation}
E_{d} = -\frac{\mu^2_s \mu_0}{4\pi a^3}\sum\limits_{i<j}\frac{
3(\hat{r}_{ij}\cdot\hat{m}_i)
(\hat{r}_{ij}\cdot \hat{m}_j)
-\hat{m}_i\cdot\hat{m}_j}{R_{ij}^3},
\label{DDI}
\end{equation}
where $\mu_s$ is the absolute value of the magnetic moment in  units of Bohr magneton, $R_{ij}=|\vec{R}_{ij}|$ are the distances between moments $i$ and $j$ in units of the lattice constant $a$, $\hat{r}_{ij}=\vec{R}_{ij}/R_{ij}$ are the unit vectors in the direction of $\vec{R}_{ij}$.

Let consider 2D square lattice of spins in the plane of $z=0$ and infinite in $x$- and $y$-directions.
The highest energy of DDI in this case corresponds to $\hat{m}_i\equiv(0,0,1)$ and the DDI energy per lattice site is
\begin{equation}
e_{d} = \frac{\mu^2_s \mu_0}{4\pi a^3}
{\sum\limits_{i=-\infty}^{+\infty}}^\prime
{\sum\limits_{j=-\infty}^{+\infty}}^\prime
\frac{
1}{\left(i^2+j^2 \right) ^{3/2}},
\label{DDI2}
\end{equation}

The sum in (\ref{DDI2}) excludes the position of the lattice site $i=j=0$ (denoted by $^\prime$) and can be calculated explicitly \cite{Borwine} 
\begin{equation}
{\sum\limits_{i=-\infty}^{+\infty}}^\prime
{\sum\limits_{j=-\infty}^{+\infty}}^\prime
\frac{1}{\left(i^2+j^2 \right) ^{3/2}}=4\cdot\beta(\frac{3}{2})\cdot\zeta(\frac{3}{2})=9.03362,
\label{DDI3}
\end{equation}
where $\beta(s)$ and $\zeta(s)$ are the values of Dirichlet beta and Riemann zeta functions for $s=\frac{3}{2}$, respectively.

Thereby, the maximum energy of DDI per one lattice site of infinite 2D square lattice of Mn/W(001) with $\mu_s=\mu_\mathrm{Mn}=3.1\mu_\mathrm{B}=179.4\times10^{-6}$ eV/T and $a=3.164$ \AA \ is 

\begin{equation}
e^{\textrm{max}}_{d} = \frac{\mu^2_s \mu_0}{4\pi a^3}\times 9.03362 = \frac{(179.4\times10^{-6}\mathrm{eV/T})^2 4\pi \times 10^{-7}\mathrm{T\!\cdot\! m/A}}{4\pi\times(3.165\times 10^{-10}\mathrm{m})^3}\times 9.03362 = 0.147 \ \mathrm{meV}.
\label{DDIfin}
\end{equation}  

Note, value in \ref{DDIfin} is the upper bound limit of DDI while the lower bound limit has negative sign, it depends on the relative orientation of the spin.
Thereby, the energy of DDI per Mn atom for Mn/W(001) system is 
\begin{equation}
-0.147 \ \mathrm{meV}\leq e_{d}\leq +0.147 \ \mathrm{meV}
\end{equation}

When upper/lower bound limit of DDI is one or two order of magnitude lower than the coupling constants of other energy terms in the system, one may neglect with the contribution of DDI. Compare to the contribution of DDI, for pairwise interactions one has to take into account the number of neighbors in particular shell.

Similar to the above approach, one can also estimate the contribution of dipole-dipole coupling between Mn and underlying Co layers.
The energy contribution of the Mn-Co interlayer DDI per one Mn spin can be written as
\begin{align}
E_{id} = &-\mu_\mathrm{Mn} \hat{m}_i \cdot \vec{H}_\mathrm{Co},
\nonumber \\
\vec{H}_\mathrm{Co}=&\frac{\mu_\textrm{Co}\mu_0}{4\pi a^3}
\sum\limits_{i=-\infty}^{+\infty}
\sum\limits_{j=-\infty}^{+\infty}\frac{
3\,\hat{r}_{ij}
(\hat{r}_{ij}\cdot \hat{m}_j)
-\hat{m}_j}{R_{ij}^3},
\label{Eid}
\end{align}
where $\vec{H}_\mathrm{Co}$ is the magnetostatic field generated by the Co layer, $\mu_\textrm{Co}\approx2\mu_\mathrm{B}$, $R_{ij}=\sqrt{i^2+j^2+k^2}$, $d_z=ka$ is the distance between Mn and Co layers. 
Under the assumption that Co layer is homogeneously magnetized $m_\textrm{Co}=(0,0,1)$ (we assume $\mu_\mathrm{Co}=2\mu_\mathrm{B}=115.8\times10^{-6}$ eV/T) and Mn atoms are situated exactly on top of the Co lattice sites, the field $\vec{H}_\mathrm{Co}$ in (\ref{Eid}) can be calculated numericaly with a high precision\cite{Mathematica}
\begin{align}
\vec{H}_\mathrm{Co}=&(0,\ 0,\ 0.019150\ \ \ \ \ \ \ \ \ \ ) \ \textrm{T},\ \ d_z=1a,\nonumber\\
\vec{H}_\mathrm{Co}=&(0,\ 0,\ 0.032455\times 10^{-3}) \ \textrm{T},\ \  d_z=2a,\nonumber \\ 
\vec{H}_\mathrm{Co}=&(0,\ 0,\ 0.060186\times 10^{-6}) \ \textrm{T},\ \  d_z=3a.\nonumber
\nonumber
\label{Hid}
\end{align}
Thereby, the energies of dipole-dipole interaction per lattice site of Mn atom, $\mu_\mathrm{Mn}=3.1\mu_\mathrm{B}=179.4\times10^{-6}$ eV/T 
\begin{eqnarray}
-0.003435\times10^{-3}\ &\textrm{meV} \leq E_{id}\leq +0.003435\times10^{-3}\ &\textrm{meV}, d_z=1a,\nonumber \\
-5.822427\times10^{-6}\ &\textrm{meV} \leq E_{id}\leq +5.822427\times10^{-6}\ &\textrm{meV}, d_z=2a,\nonumber \\ 
-10.79737\times10^{-9}\ &\textrm{meV} \leq E_{id}\leq +10.79737\times10^{-9}\ &\textrm{meV}, d_z=3a,\nonumber
\label{Eid2}
\end{eqnarray}
where the sign of energy contribution depends on the sign of $m_z$ of Mn atom with respect to $\vec{H}_\mathrm{Co}$.
The field $\vec{H}_\mathrm{Co}$ decays exponentially with the distance $d_z$ and in this particular case for $d_z \geq 2a$ the dipole-dipole coupling between the Mn and Co layers can be neglected.
 
\section{Details of the Monte Carlo calculations}

The MC simulation \cite{MCS} has been used to define the equilibrium parameters of spin structures, phase transitions and elliptical instability as well as collapse of isolated skyrmion (iSk) discussed in the main text.
We do the simulation of annealing process to identify the ground state of the system at different magnetic field and phase transition between ordered and paramagnetic states \cite{Laarhoven}.
Phase transition lines of the first order ($B_\textrm{tr1}$ and $B_\textrm{tr2}$) have been found by the calculation of the free energy difference. 
We have used the Bennett acceptance ratio method \cite{Bennett76}, which is based on the counting of transition probabilities between competing phases:
\begin{equation}
\Delta F_\textrm{AB} = -\frac{1}{k_{B}T}\ln{\frac{\rho(\textrm{B} \rightarrow \textrm{A})}{\rho(\textrm{A} \rightarrow \textrm{B})}},
\label{freeEnergy}
\end{equation}
where $\rho(\textrm{B} \rightarrow \textrm{A})$ ($\rho(\textrm{A} \rightarrow \textrm{B})$) denote the transition probability from phase B (A) to phase A (B). 
Free energy difference is estimated after every 10 MC steps. 
Calculations are carried out till the saturation of the free energy difference $\Delta F_\textrm{AB}$, within at least 10$^\textrm{6}$ MC steps. 
Calculation has been done for fixed temperature and varying magnetic field with the step $\Delta B = 0.25$ T.

Dependence of the equilibrium period of the SS, period of the skyrmion lattice (SkL) and equilibrium magnetization on applied magnetic field has been found as following. 
By choosing finite size of domains to fit spin structure of particular periodicity, we define the range of field where energy density of the state with corresponding periodicity has a lowest value. 
In field dependence of iSk has been obtained by full relaxation for simulated domain of $64 \times 64$ spins size at each value of applied magnetic field.

Topological lability lines $B_\textrm{SSL}$ and $B_\textrm{SkLL}$ and lines confining the area of stable iSk $B_\textrm{iSkE}$ and $B_\textrm{iSkC}$ are found by long time relaxation within at least 5$\times$10$^\textrm{6}$ MC steps at each fixed field following with interval of $\Delta\textrm{B}$ = 0.1 T, at each fixed temperature.
Instantaneous and average values of topological charge (TC), magnetization and internal energy of the system are obtain with the snapshots of the system collected after each 100 MC step. 
At each temperature step with $\Delta T = 1$K we use $10^4$ MC steps for relaxation and $10^5$ MC steps for statistic.
A typical size of the simulated domain in our MC simulations is of about 128 $\times$ 128. 

Instead of classical Metropolis algorithm, we use \textit{small step algorithm} \cite{Nowak} together with simple feedback algorithm keeping value of rejection-acceptation ration at about 0.5.
This allows us increase in efficiency of MC sampling by approximately one order of magnitude without any lost of precision. 
Nevertheless, due to the large number of energy terms as well as interacting shells, the calculation of the total energy is extremely time consuming. 
To increase the efficiency of the calculations we use massively parallel computation within Graphic Processing Unit instead of Central Processing Unit, allowed by recently developed CUDA technology. 
After optimizing the MC algorithm used with CUDA-Fortran (extension of standard Fortran language for GPU programming), we achieved efficiency of GPU version of the programming code approximately 20 times higher compare to CPU Fortran version.

\section{Details of the B-T phase diagram.}

\begin{figure}[!t]
 \centering
\includegraphics[width=8.5cm]{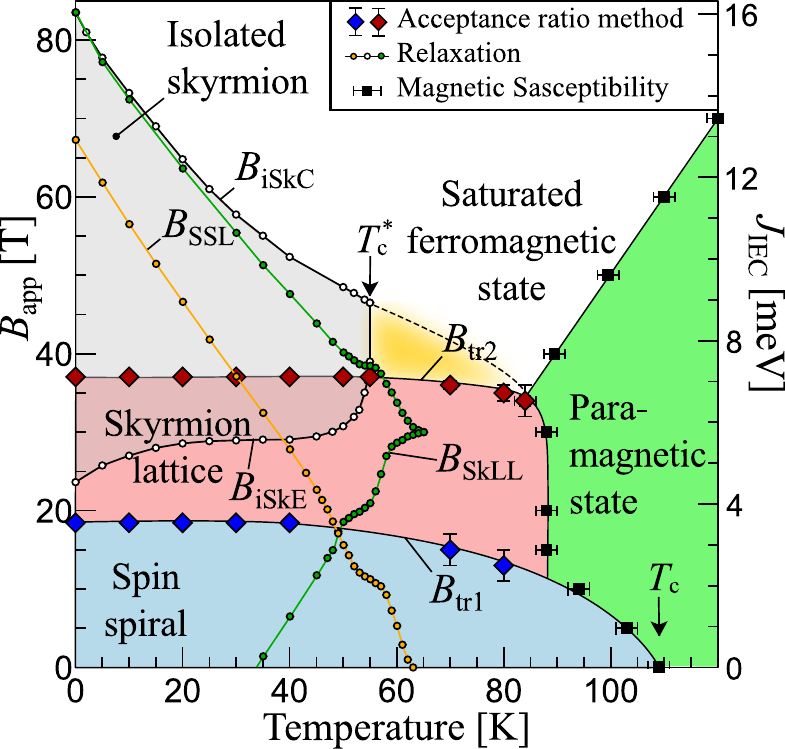}
 \caption{ Details of the phase diagram including topological lability lines.}
 \label{fig_PD-detail}
\end{figure}

\begin{figure*}[!t]
\includegraphics[width=12.5cm]{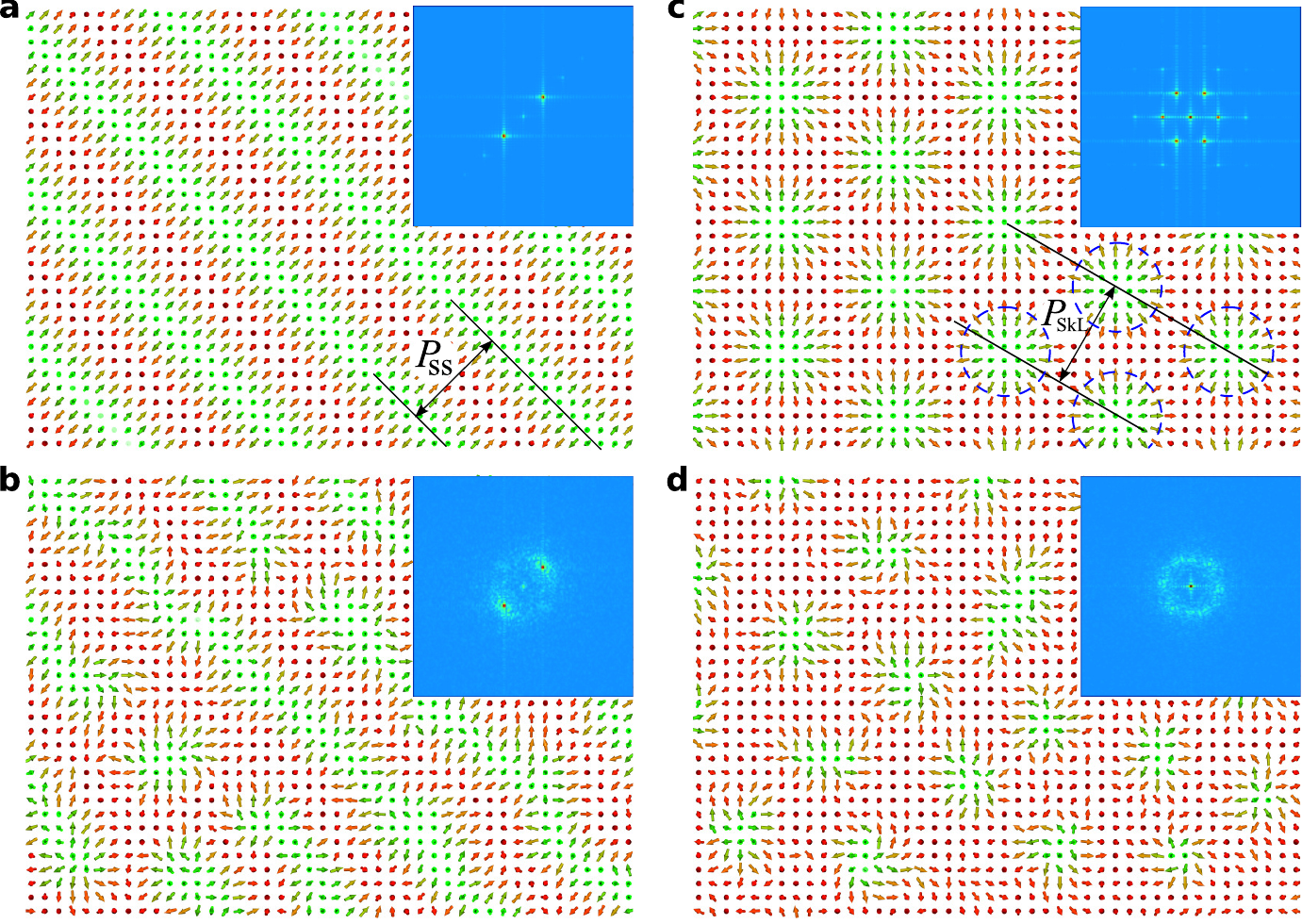}\\
 \caption{ Snapshot of the system corresponding to (a) an ideal SS taken at applied field $B = 10$ T and $T = 0$ K, (b) the distorted SS state due to the thermal fluctuations taken at $B = 10$ T and $T=75$ K, (c) an ideal hexagonal lattice of skyrmions taken at $B = 30$ T and $T=0$ K, and (d) the skyrmion lattice which lost the long range order due to the thermal fluctuations taken at $B = 30$ T and $T=75$ K.
Insets show the Fourier transformation of the out-of-plane magnetization of the corresponding spin structure.
The square cell in Fourier transformation image corresponds to the BZ related to the two-dimensional atomic lattice.}
 \label{fig_SS_SL}
\end{figure*}

\begin{figure}[!hb]
 \centering
\includegraphics[width=6.5cm]{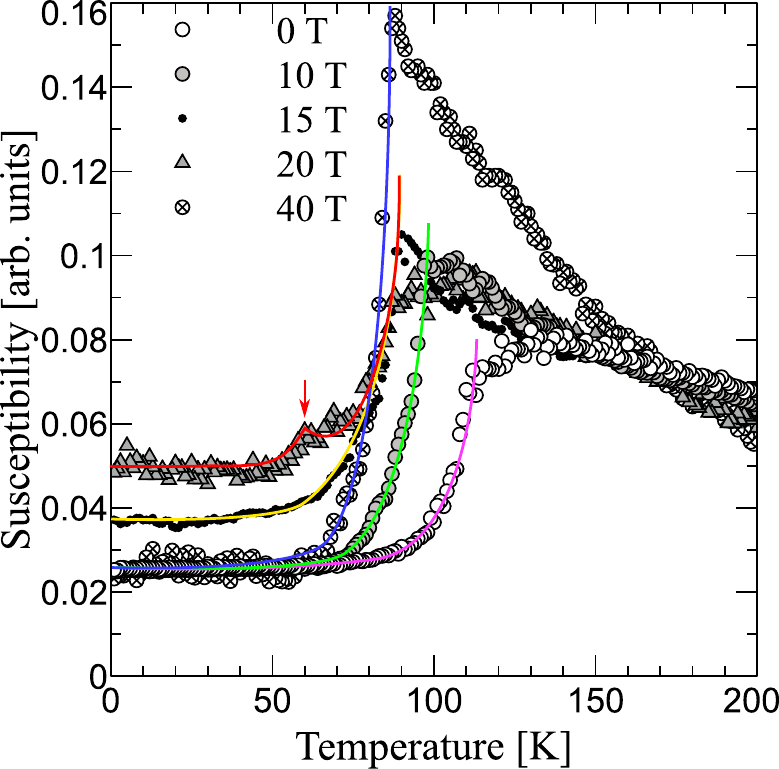}
 \caption{ Temperature dependence of magnetic susceptibility at different applied magnetic field. The solid lines follow the eye. Note an additional peak marked with red arrow for applied filed equal 20 T, it can be attributed with the order-disorder transition of SkL, see critical line $B_\textrm{SkLL}$ in phase diagram, Fig.~\ref{fig_PD-detail}.}
 \label{fig_susceptibility}
\end{figure}

\begin{figure}[!b]
\includegraphics[width=6.5cm]{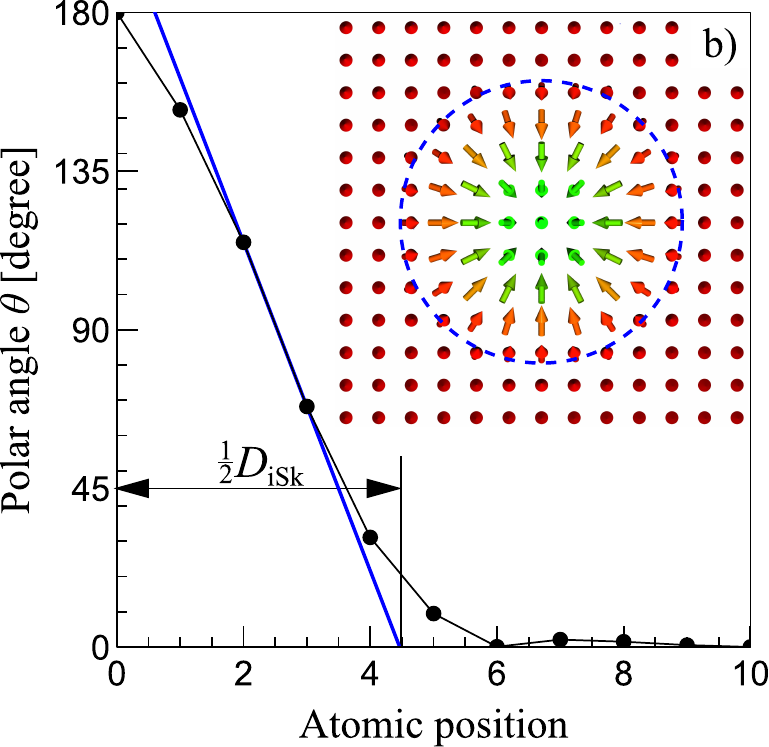}\\
\caption{ The size of an iSk.
Polar angle of the spins on atomic positions along the line crossing the center of iSk (solid circles).
Solid blue line is the tangent to the skyrmion profile at its half width.
The tangent line fitted as linear function $y = b - a*|x|$ connects two points which are nearest to the value of $\theta=90^{\circ}$.
Diameter of the skyrmion is defined as $D_\textrm{iSk} = 2b/a$.
Inset shows the central part of the simulated domain ($31 \times 31$ spins) with isolated domain corresponding to the profile in the main figure.
The dashed circle denote the size of the skyrmion. Shown iSk corresponds to the state at applied field of $B = 25$ T.
Presented approach for estimation of iSk size is similar to approach used by Bogdanov and Hubert \cite{Bogdanov_94_pss} for continuum micromagnetic model and can be considered as an analogue for the case of discrete lattice.}
\label{fig_iSk_size}
\end{figure}

Figure \ref{fig_PD-detail} includes more details about the magnetic phase diagram compare to that presented in the main text for Mn/W(001) system.
Below the critical magnetic field $B_\textrm{tr1}$, the SS state which is the lowest energy state is shown in Fig.~\ref{fig_SS_SL}a.
The Fourier transformation of the out-of-plane magnetization in the inset shows two pronounced peaks corresponding to the periodicity of the homogeneous spiral.
Within the range of field between $B_\textrm{tr1}$ and $B_\textrm{tr2}$, the lowest energy state corresponds to the hexagonal lattice of skyrmions as presented in Fig.~\ref{fig_SS_SL}c.
The six pronounced peaks surrounding the central peak in the Fourier transformation corresponds to the hexagonal symmetry of the SkL.
$P_\textrm{SS}$ and $P_\textrm{SkL}$ are the periods of the homogeneous SS and hexagonal SkL, respectively.
Above $B_\textrm{tr2}$ saturated ferromagnetic state has the lowest energy.

In the phase diagram Fig.~\ref{fig_PD-detail}, we have added the topological lability lines for the SS and hexagonal SkL states.
Here, the topological lability of a magnetic state defines the range of applied magnetic field and temperature outside of which the TC of the stable or metastable state is no longer conserved.
We follow the definition of TC on a discrete lattice given by Berg and L\"uscher \cite{Berg81}.
Since, the period of SkL depends on applied magnetic field (see the main text), the TC density defined as the number of skyrmions for a fixed number of spins is also field dependent.
The TC density remains conserved within the range of applied magnetic field bounded by the critical field line $B_\textrm{SkLL}$, outside of it the conservation of TC is violated due to the elliptical instability and/or collapse of skyrmions.
This may result in either SS state (low field case) or homogeneous ferromagnetic state (high field case).
Outside the lability lines within the SkL phase the long range order of the hexagonal phase is lost due to the thermal fluctuations but the average distance between skyrmions remains conserved, see Fig.~\ref{fig_SS_SL}d and inset.
Below the lability line $B_\textrm{SSL}$, the SS state has zero TC while at and above it, the thermal fluctuation results in defects, see the spin structure in Fig.~\ref{fig_SS_SL}b,which lead to nonzero TC.
Such an instability of SS and SkL driven by thermal fluctuations leads to the loss of the long range order, while the short range order is still conserved. Such distorted states also can be interpreted as so-called fluctuation-disordered state \cite{Janoschek_PRB,Rozsa_PRB}.

Transition temperatures between the paramagnetic and the ordered phases are determined from the magnetic susceptibility-peak position, see Fig.~\ref{fig_susceptibility}.
It is worth to mention that the topological lability lines discussed above can be also identified as the transition lines between ordered and disordered statesi, which is reflected by a small additional peak in the temperature dependence of magnetic susceptibility, see red arrow in Fig.~\ref{fig_susceptibility}.

The diameter of an iSk $D_\textrm{iSk}$ in the main text is determined according to Fig.~\ref{fig_iSk_size}.

\section{Zero field isolated skyrmion stabilized by anisotropy.}

\begin{figure*}[!h]
 \centering
   \includegraphics[width=0.90\textwidth]{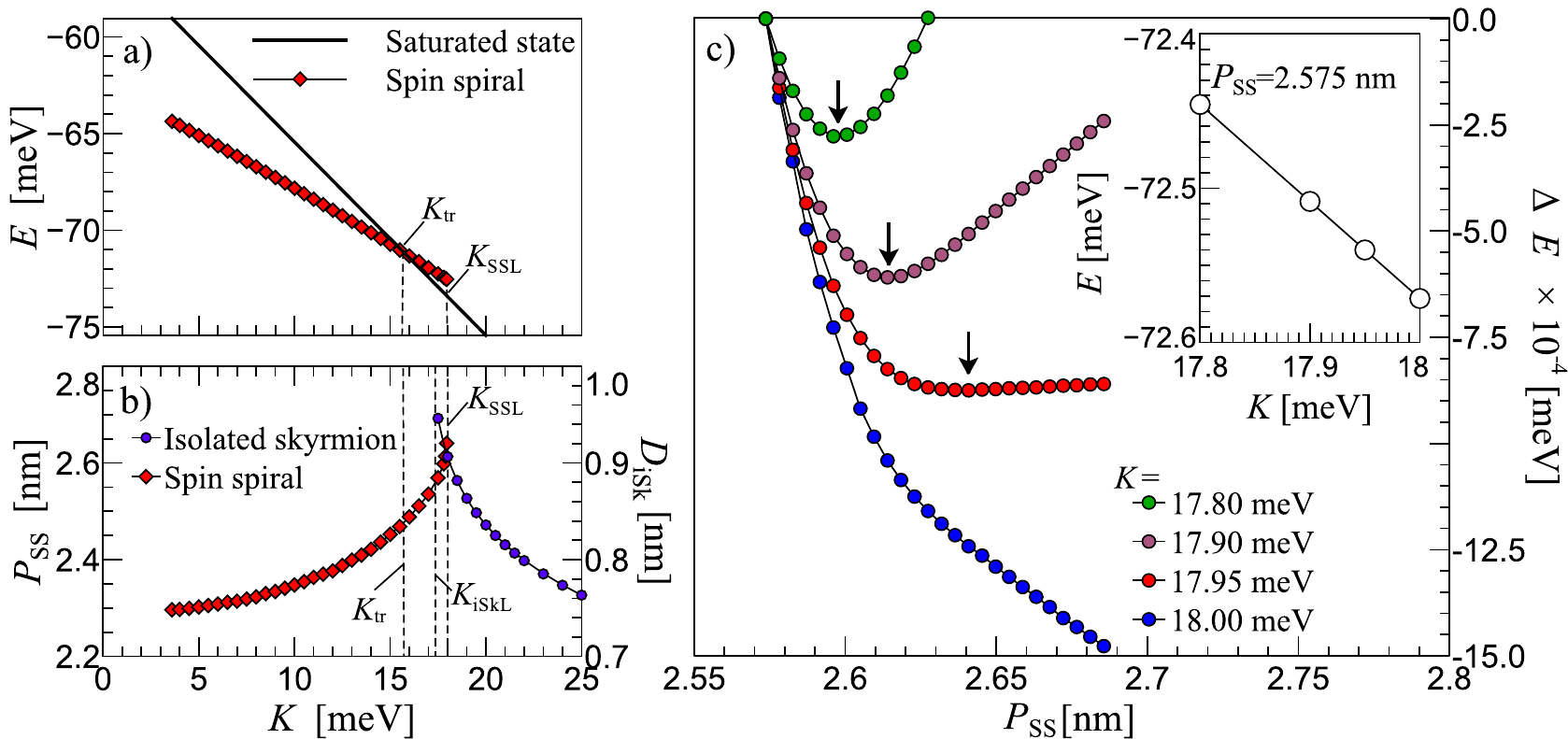}
 \caption{ (a) Energy density of SS state and saturated ferromagnetic state calculated at zero magnetic field for varying value of uniaxial anisotropy \textit{K}, with fixed values of other
coupling constants corresponding to the model of Mn/W(001). $K_\textrm{tr}=15.7$ meV is the anisotropy value corresponding to the phase transition between SS and saturated state.
(b) Dependence of the equilibrium period of the spiral $P_\textrm{SS}$ and diameter of the iSk $D_\textrm{iSk}$ with respect to anisotropy at zero applied field. 
The maximum anisotropy value $K_\textrm{SSL}=17.95$ corresponding to the stability of SS is slightly higher than that $K_\textrm{iSkL}=17.35$ for stabilizing iSk. 
Note a different scale range for $P_\textrm{SS}$ on the left axis and for $D_\textrm{iSk}$ on the right axis. 
(c) Energy profiles with respect to spiral period $P_\textrm{SS}$ for different values of anisotropy near the stability point $K_\textrm{SSL}$. 
Arrows denote the position of the minimum corresponding to the equilibrium period. 
All energies are counted as difference $\Delta E= E(P^*_\textrm{SS})-E(P_\textrm{SS})$, where $P^*_\textrm{SS}=2.575$ nm is a value of the SS period arbitrarily chosen to have a reasonable scale 
for all profiles. Inset shows dependence of the energy at fixed period $P^*_\textrm{SS}$ and different values of anisotropy.}
\label{fig_K}
\end{figure*}

We have estimated the critical anisotropy required to stabilize the saturated state for the case of Mn/W(001). 
In Fig.~\ref{fig_K}a we have shown the energy density dependence for the equilibrium SS and saturated states. 
The transition between the states corresponds to $K_\textrm{tr}\simeq 15.7$ meV, which is more than four times larger than the calculated value for Mn/W(001). 
Spin spiral may exist as a metastable state within a range $K_\textrm{tr} < K < K_\textrm{SSL}$, where $K_\textrm{SSL}$ corresponds to the anisotropy above which SS loses the stability, lability point.
Figure \ref{fig_K}c illustrates the energy profile corresponding to the SS state for different anisotropy values near the lability point $K_\textrm{SSL} = 17.95$ meV. 
For $K > K_\textrm{SSL}$ the only solution is $P_\textrm{SS}\rightarrow \infty$ which corresponds to the saturated state.

Isolated skyrmions exist as a metastable state for $K > K_\textrm{iSkL} > K_\textrm{tr}$, where $K_\textrm{iSkL}$ is a stability point for iSk. 
In this region, the size of skyrmions is extremely small, $D_\textrm{iSk} < 1$ nm compare to the smallest skyrmion which we found under applied magnetic field, see main text. 


\end{document}